\documentclass[showpacs, reprint,
nofootinbib,
amsmath, amssymb,
aps,
]{revtex4-1}

\usepackage{graphicx}% Include figure files
\usepackage{dcolumn}% Align table columns on decimal point
\usepackage{tabu}
\usepackage{bm}% bold math
\usepackage{lipsum}
\usepackage[bookmarks=false]{hyperref}
\usepackage{hyperref}% add hypertext capabilities
\hypersetup{pdfstartview={XYZ null null 1.00}}
\begin{document}

\preprint{APS/123-QED}

\title{Density perturbations in $f(R,\phi)$-gravity with an application to the (varying power)-law model}

\author{Fay\c{c}al Hammad}
\email{fhammad@ubishops.ca}
\affiliation{Physics Department \& STAR Research Cluster, Bishop's University,\\
2600 College St., Sherbrooke, QC, J1M 1Z7, Canada\\
Physics Department, Champlain College-Lennoxville,\\
2580 College St., Sherbrooke, QC J1M 0C8, Canada}


\begin{abstract}
Density perturbations in the cosmic microwave background within general $f(R,\phi)$ models of gravity are investigated. The general dynamical equations for the tensor and scalar modes in any $f(R,\phi)$ gravity model are derived. An application of the equations to the (varying power)-law modified gravity toy-model is then made. Formulas and numerical values for the tensor-to-scalar ratio, the scalar tilt and the tensor tilt are all obtained within this specific model. While the model cannot provide a theoretical reason for the value of the energy scale at which inflation should occur, it is found, based on the latest observations of the density perturbations in the sky, that the model requires inflation to occur at an energy scale less than the GUT-scale; namely, $\sim10^{14}\,{\rm GeV}$. The different energy intervals examined here show that the density perturbations recently obtained from observations are recovered naturally, with very high precision, and without fine tuning the model's parameters.
\end{abstract}

\pacs {98.80.-k, 98.80.Es, 98.80.Cq}
%PACS, the Physics and Astronomy Classification Scheme.
\maketitle

\section{Introduction}\label{sec:1}
The increased precision of recent observations of the cosmic microwave background (CMB)~\cite{Planck2013, Planck2015} gives a precious tool to put to test any model of gravity that departs from general relativity. The simplest way general relativity could make a signature in the CMB is through a minimal coupling of gravity with the scalar field responsible for inflation \cite{Guth} (see also Refs.~\cite{Sato,Linde,Albrecht} and \cite{Starobinsky,Fang} for the early works on inflation). It is well known, however, that many of the inflationary models based on a single field require unsatisfactory fine-tuning of their parameters (see e.g. Ref.~\cite{Martin} for a nice survey.) This is, it turns out, one of the reasons that make modified gravity models attractive because less fine-tuning is required in these models and one can use them to study cosmology \cite{Nojiri,StarobinskyCosmo,Sawicki} and density perturbations during matter domination \cite{Carroll*,Bean*,Song1*,Hu*,Song2*,Amendola*,Tsujikawa1*,Tsujikawa2*,Pogosian*,Cruz-Dombriz*,Motohashi*,Koyama*,Gannouji*,Tsujikawa3*,Narikawa*,Matsumoto*,Zhang*,Li*}.

The advantage of modified gravity over general relativity minimally coupled to a scalar field is provided by the higher powers of curvature that take the lead  at small scales. The simplest modified gravity model known to accommodate inflation is Starobinsky's power-law model \cite{Starobinsky}. In this model the Hilbert-Einstein action is increased by a term containing the second power of the Ricci scalar.

Many more power-law models have been introduced since then, either suggested, along with other higher-order curvature invariants, by low-energy effective Lagrangians in string theory and other approaches to quantum gravity \cite{Nojiri+*+}, or introduced just for phenomenological reasons. The latter are more or less favored among themselves for various cosmological reasons \cite{Faulkner, Clifton, MGInflationSurvey2015}, such as, the solar-system tests \cite{Hu*,Guo}, the weak lensing \cite{Higuchi**,Tsujikawa**}, the cluster abundance \cite{Schmidt} and the baryon acoustic oscillations \cite{Yang,Lombriser} (see also Ref.~\cite{Hammad1} for a link between modified gravity and microscopic physics.) In these more complicated power-law models, known collectively as $f(R)$-gravity models, one finds one or more powers of the Ricci scalar in the gravitational action. In order to make the models agree with the various constraints from observations, some of them require integer powers of the Ricci scalar while others just require that the power be a real number (see e.g. Refs.~\cite{Amendola*,Tsujikawa**,Nojiri**+,Capozziello**+,Nojiri+**}.)

This \textit{a priori} continuum of possibilities for the powers of the Ricci scalar that might appear in the additional term inside the Hilbert-Einstein action is what has motivated the introduction of the (varying power)-law modified gravity model in Ref.~\cite{Hammad2}. Indeed, instead of choosing a given fixed power for the Ricci scalar in the action, one might just leave the power to vary, as a free additional degree of freedom, in the hope that it adjusts itself according to the physical environment it finds itself in. The power of the Ricci scalar thus becomes a scalar field governed by precise dynamics just as any other ordinary scalar field. It has been pointed out in Ref.~\cite{Hammad2} that the model can easily be used to generate inflation. Moreover, it has been shown in Ref.~\cite{Hammad3} that this highly non-linear coupling of the scalar field with gravity protects the field from any fifth force test due to the huge mass the field acquires within any low-curvature environment, and this for whatever initial mass the field happens to start with. This model belongs to the category of modified gravity models known as $f(R,\phi)$ models.

In fact, besides being a source of the possible early inflationary expansion of the Universe, a scalar field has also found applications in modified gravity models. A scalar field non-minimally coupled to gravity is mostly found in string theories where it plays the role of a dilaton field \cite{StringBook}. Just as with $f(R)$ models of gravity, more general models combining one or many scalar fields with the curvature tensors have been studied in the literature under the name of scalar-tensor theories of gravity \cite{MGReview}. The simplest and most well known are those combining only one scalar field $\phi$ and the Ricci scalar $R$ inside a gravitational action whose Lagrangian is a functional $f(R,\phi)$ of these two scalars. In these models, one combines the advantages of having higher-order curvature terms with those of having an additional degree of freedom for spacetime. These models have been extensively studied in the literature with applications in cosmology in general \cite{Boisseau***,Esposito***,Riazuelo***,Acquaviva***} and in the study of density perturbations during matter domination \cite{Gannouji+,Song+}.

The aim of the present paper is to first derive the general dynamical equations for the tensor and scalar perturbation modes during inflation in a general $f(R,\phi)$ model of gravity. To the best of our knowledge, this specific analysis has not been given elsewhere. The general theory of density perturbations in $f(R,\phi)$-gravity has been exposed in Ref.~\cite{Felice-}. However, the focus in that reference was not on the perturbations relevant for inflation, and hence no derivation is given for the dynamical equations of the tensor and scalar modes useful for inflation.

The second goal of this paper is then to use these results for general $f(R,\phi)$ models to examine the consequences of the highly non-linear coupling in the (varying power)-law modified gravity on the density perturbations in the CMB. Indeed, as pointed out in Ref.~\cite{Hammad2}, because of its high non-linearity, this model cannot be reduced to an $f(R)$ model as usual $f(R,\phi)$ models do by a suitable redefinition of the scalar field \cite{MGReview}. The remarkable result is that the model is able to reproduce at very high accuracy some of the observed values in the sky without making any fine-tuning on its parameters.

Just like in single-field inflationary models, this model allows a very slowly decreasing Hubble parameter during inflation and, hence \cite{Mukhanov*}, a small, but non-vanishing red tilt of the spectrum of the curvature perturbations, results. Furthermore, the model predicts, just as it is required from inflation \cite{Starobinsky*}, a nearly scale-invariant power spectrum of the primordial tensor perturbations. The fact that the scalar field controls the power of the Ricci scalar is what makes the scalar perturbations that couple to matter energy density and pressure perturbations dominate the tensor modes which exist even in the absence of matter in the form of gravitational waves.

As we shall see, although the model might only be phenomenological, all it requires as input is the energy scale at which one decides to apply it. In this paper we choose the scale to be the scale of inflation itself, i.e., less than the grand unified theories (GUT) scale, as suggested by the model when confronted with the latest observations \cite{Planck2015}.

The outline of the reminder of this paper is as follows. In Section~\ref{sec:2}, we derive the general dynamical equations of the tensor and the scalar perturbations within a general $f(R,\phi)$-gravity model. In Section~\ref{sec:3}, we recall the (varying power)-law model's action from which one extracts the scalar field's potential, the Klein-Gordon equation for the scalar field, and the modified Friedmann equation for the Hubble parameter. We then deduce the approximations corresponding to these three equations relevant for the early inflationary expansion of the Universe. Using these approximate equations, we compute the numerical values of the analogue of the slow-roll parameters in usual single-field inflationary models. These numerical values of the parameters are used in subsections \ref{VPLScalarModes} and \ref{VPLTensorModes}, along with the dynamical equations, to study the density perturbations implied by the model. The tensor-to-scalar ratio, the tensor and the scalar tilts are evaluated and compared with observations. We end this paper with a discussion and conclusion section.

\section{Scalar vs. tensor perturbations in $f(R,\phi)$ gravity}\label{sec:2}
In this section we shall derive the various equations needed to study the primordial density perturbations in any $f(R,\phi)$ model of gravity. In single-field inflationary models, the power spectrum $\mathcal{P}_{h}$ of the tensor perturbations and the power spectrum $\mathcal{P}_{\mathcal{R}}$ of the scalar perturbations in a de Sitter or quasi-de Sitter inflation are given by very simple expressions featuring the Hubble parameter $H$ and the slow-roll parameters of the model (see e.g. Ref.~\cite{Liddle}). In more general $f(R,\phi)$ models of gravity, however, the field equations of the theory are more complicated \cite{Felice++}, and hence one does not expect \textit{a priori} to find such simple formulae. In what follows, we shall therefore first derive the general dynamical equations of the scalar and tensor perturbations in any $f(R,\phi)$ model of gravity. These equations will allow one to extract the corresponding power spectra formulae from which one can deduce the expressions of the tensor-to-scalar ratio, the scalar tilt and the tensor tilt for any specific $f(R,\phi)$ model.

\subsection{Scalar modes in $f(R,\phi)$ gravity}\label{TSScalarModes}
The field equations of a general $f(R,\phi)$-gravity model are obtained by varying the action, $M_{P}^2/(16\pi)\int\mathrm{d}^4x\sqrt{g}[ f(R,\phi)-(\partial\phi)^2-m^2\phi^2]$ with respect to the metric $g_{\mu\nu}$ \cite{MGReview,Felice++}. Here, $f(R,\phi)$ is a regular functional of the Ricci scalar $R$ and the scalar field $\phi$, and $M_{P}$ is the Planck mass while $m$ is the scalar field's mass. We shall work throughout this paper with the natural units $\hbar=c=1$. The metric field equations then read \cite{MGReview,Felice++},
\begin{equation}\label{FieldEquations}
f_{,R}R_{\mu\nu}-\frac{1}{2}g_{\mu\nu}f+(g_{\mu\nu}\Box-\nabla_{\mu}\nabla_{\nu})f_{,R}=T_{\mu\nu}^{\phi},
\end{equation}
where $T_{\mu\nu}^{\phi}=\partial_{\mu}\phi\partial_{\nu}\phi-\frac{1}{2}g_{\mu\nu}(\partial\phi)^{2}-\frac{1}{2}g_{\mu\nu}m^{2}\phi^{2}$ is the energy-momentum tensor of the scalar field $\phi$. We will keep neglecting henceforth $T^{m}_{\mu\nu}$, the energy-momentum tensor of ordinary matter fields, as it is a natural assumption to discard the contribution of ordinary matter during the inflationary epoch. We will also neglect the mass term of the scalar field $\phi$ as it is irrelevant at these scales, in $f(R,\phi)$ models in general, and in the (varying power)-law model in particular \cite{Hammad2, Hammad3}. In fact, besides the non-minimally coupled potential of the scalar field, already taken into account inside the $f(R,\phi)$-term, the dominant contribution of $\phi$ comes mainly from its kinetic energy $\frac{1}{2}\dot{\phi}^{2}$.

The field equations (\ref{FieldEquations}) describe the dynamics of the unperturbed spacetime background. We proceed now to the introduction of perturbations in the latter. First, by considering a small perturbation $|h_{\mu\nu}|\ll1$ of the background Friedmann-Lema\^{\i}tre-Robertson-Walker (FLRW) metric $\bar{g}_{\mu\nu}=\mathrm{diag}(-1,a^{2},a^{2},a^{2})$, where $a(t)$ is the positive scale factor, such that $g_{\mu\nu}=\bar{g}_{\mu\nu}+h_{\mu\nu}$, we can extract from (\ref{FieldEquations}) the resulting perturbed field equations as follows,
\begin{multline}\label{PerturbedFE}
f_{,R}\delta R_{\mu\nu}+(\bar{R}_{\mu\nu}-\nabla_{\mu}\nabla_{\nu})\delta f_{,R}-\frac{1}{2}\bar{g}_{\mu\nu}\delta(f-2\Box f_{,R})\\-\frac{1}{2}h_{\mu\nu}(f-2\Box f_{,R})+\delta\Gamma_{\mu\nu}^{\lambda}\nabla_{\lambda}f_{,R}=\delta T^{\phi}_{\mu\nu}.\vspace*{-0.9cm}
\end{multline}
We followed here the convention used in Ref.~\cite{Weinberg} by distinguishing the unperturbed background quantities with an overbar. A direct computation gives the non-vanishing Ricci tensor components of the background spacetime \cite{Weinberg}, $\bar{R}_{00}=-3(\dot{H}+H^{2})$, $\bar{R}_{0i}=0$, and $\bar{R}_{ij}=a^{2}\delta_{ij}(\dot{H}+3H^{2})$. For the Christoffel symbols, the relevant non-vanishing components are \cite{Weinberg}, $\bar{\Gamma}^{i}_{0j}=H\delta_{j}^{i}$ and $\bar{\Gamma}^{0}_{ij}=a^{2}H\delta_{ij}$.

Next, in order to study the scalar perturbations, one usually chooses the perturbed FLRW metric written in the Newtonian (or longitudinal) gauge \cite{Weinberg} (see also Refs.~\cite{Mukhanov,Riotto}),
\begin{equation}\label{NewtonianGauge}
\mathrm{d}s^{2}=-(1+2\Phi)\mathrm{d}t^{2}+a^{2}(t)(1-2\Psi)\delta_{ij}\mathrm{d}x^{i}\mathrm{d}x^{j}.
\end{equation}
The small dimensionless quantities $\Phi$ and $\Psi$, called the Newtonian potential and spatial curvature, respectively, give the perturbations $h_{\mu\nu}$ of the metric about its background value $\bar{g}_{\mu\nu}$. Thus, the perturbed components of the connection that will be relevant for us are, $\delta\Gamma^{0}_{0i}=\partial_{i}\Phi$, $\delta\Gamma^{0}_{00}=\dot{\Phi}$ and $\delta\Gamma_{ij}^{0}=-a^{2}\delta_{ij}(2H\Phi+2H\Psi+\dot{\Psi})$. On the other hand, with such a metric as (\ref{NewtonianGauge}), the perturbed components of the scalar field's energy-momentum tensor read, $\delta T_{00}^{\phi}=\dot{\phi}\delta\dot{\phi}$, $\delta T_{0i}^{\phi}=\dot{\phi}\partial_{i}\delta\phi$ and $\delta T_{ij}^{\phi}=a^{2}\delta_{ij}[\dot{\phi}\delta\dot{\phi}-(\Psi+\Phi)\dot{\phi}^{2}]$, and for the perturbed Ricci components, we easily compute,
\begin{equation}\label{scalardeltaR0i}
\delta R_{0i}=2\partial_{i}(H\Phi+\dot{\Psi}),
\end{equation}
\vspace{-0.5cm}
\begin{equation}\label{scalardeltaR00}
\delta R_{00}=3\ddot{\Psi}+6H\dot{\Psi}+3H\dot{\Phi}+\frac{\nabla^{2}}{a^{2}}\Phi,
\end{equation}
\vspace{-0.5cm}
\begin{multline}\label{scalardeltaRij}
\delta R_{ij}=\nabla_{i}\nabla_{j}(\Psi-\Phi)-a^{2}\delta_{ij}\Bigg[\ddot{\Psi}+5H\dot{\Psi}
\\+\left(2\dot{H}+6H^{2}-\frac{\nabla^{2}}{a^{2}}\right)\Psi+2H\dot{\Phi}+(2\dot{H}+6H^{2})\Phi\Bigg].
\end{multline}

First, by plugging the components (\ref{scalardeltaRij}) of the perturbed Ricci tensor inside the perturbed field equations (\ref{PerturbedFE}), we can isolate the following $ij$-components, with $i\neq j$:
\begin{equation}
f_{,R}\partial_{i}\partial_{j}(\Psi-\Phi)-\partial_{i}\partial_{j}\delta f_{,R}=0,
\end{equation}
which integrate to give, $\delta f_{,R}=f_{,R}(\Psi-\Phi)$. On the other hand, by plugging the $0i$-components (\ref{scalardeltaR0i}) of the perturbed Ricci tensor inside the perturbed field equations (\ref{PerturbedFE}), we obtain,
\begin{equation}\label{scalar0i}
2f_{,R}(H\partial_{i}\Phi+\partial_{i}\dot{\Psi})-\partial_{0}\partial_{i}\delta f_{,R}+H\partial_{i}\delta f_{,R}+\partial_{i}\Phi\dot{f}_{,R}=\dot{\phi}\partial_{i}\delta\phi.
\end{equation}
When integrating these equations with respect to the spatial coordinates $i$, and then substituting $\delta f_{,R}$ in terms of $\Psi$ and $\Phi$, we find the following single differential equation:
\begin{equation}\label{simplescalar0i}
f_{,R}(\dot{\Phi}+\dot{\Psi})+(Hf_{,R}-\dot{f}_{,R})(\Phi+\Psi)+3\dot{f}_{,R}\Phi=\dot{\phi}\delta\phi.
\end{equation}
By differentiating this equation once with respect to the time coordinate and then making use of (\ref{simplescalar0i}) once more inside the resulting equation, we arrive at the following second-order differential equation in time $t$, that we are going to make use of later:
\begin{multline}\label{intermediatescalar0i}
\ddot{\Phi}+\ddot{\Psi}+\left(H-\frac{\ddot{\phi}}{\dot{\phi}}\right)(\dot{\Phi}+\dot{\Psi})
\\+\left[\dot{H}+H\left(\frac{\dot{f}_{,R}}{f_{,R}}-\frac{\ddot{\phi}}{\dot{\phi}}\right)
-\frac{\ddot{f}_{,R}}{f_{,R}}+\frac{\dot{f}_{,R}}{f_{,R}}\frac{\ddot{\phi}}{\dot{\phi}}\right](\Phi+\Psi)\\
+\frac{3\dot{f}_{,R}}{f_{,R}}\dot{\Phi}
+3\left(\frac{\ddot{f}_{,R}}{f_{,R}}-\frac{\dot{f}_{,R}}{f_{,R}}\frac{\ddot{\phi}}{\dot{\phi}}\right)\Phi=\frac{\dot{\phi}\delta\dot{\phi}}{f_{,R}}.
\end{multline}

Let us now write explicitly the $00$-component as well as the $ij$-components of the perturbed field equations (\ref{PerturbedFE}). We have the following equations, where for the $ij$-components we divide both sides of the equalities by $a^{2}$:
\begin{multline}\label{Scalar00}
f_{,R}\delta R_{00}-(3\dot{H}+3H^{2}+\partial_{0}^{2})\delta f_{,R}+\frac{1}{2}\delta(f-2\Box f_{,R})\\
+\Phi(f-2\Box f_{,R})+\dot{\Phi}\dot{f}_{,R}=\dot{\phi}\delta\dot{\phi}.
\end{multline}
\vspace{-0.9cm}
\begin{multline}\label{Scalarij}
\frac{f_{,R}}{a^{2}}\delta R_{ij}-\frac{\nabla_{i}\nabla_{j}}{a^{2}}\delta f_{,R}
\\+\delta_{ij}\!\left[(\dot{H}+3H^{2})\delta f_{,R}-\frac{1}{2}\delta(f-2\Box f_{,R})+\Psi(f-2\Box f_{,R})\right]
\\-\delta_{ij}(2H\Phi+2H\Psi+\dot{\Psi})\dot{f}_{,R}+\delta_{ij}(\Psi+\Phi)\dot{\phi}^{2}=\delta_{ij}\dot{\phi}\delta\dot{\phi}.
\end{multline}
By adding (\ref{Scalar00}) and (\ref{Scalarij}), after multiplying the first by $\delta_{ij}$, the term $\delta(f-2\Box f_{,R})$ cancels out. Also, by using the identity, $f_{,R}(\Psi-\Phi)=\delta f_{,R}$, the term $\nabla_{i}\nabla_{j}\delta f_{,R}$ in (\ref{Scalarij}) cancels out with a similar term provided by the presence of $\delta R_{ij}$ in that equation, as it follows from (\ref{scalardeltaRij}). Then, after substituting the full expressions (\ref{scalardeltaR00}) and (\ref{scalardeltaRij}) of $\delta R_{00}$ and $\delta R_{ij}$, respectively, and rearranging the various terms, the sum of (\ref{Scalar00}) and (\ref{Scalarij}) yields,
\begin{multline}\label{2phideltaphi}
\!\!\!\!\!\!\ddot{\Phi}+\ddot{\Psi}+\left(H-\frac{3\dot{f}_{,R}}{f_{,R}}\right)(\dot{\Psi}+\dot{\Phi})
+\left(4\dot{H}+2\frac{\ddot{f}_{,R}}{f_{,R}}\right)\Phi+\frac{6\dot{f}_{,R}}{f_{,R}}\dot{\Phi}
\\-\left(4\dot{H}+6H^{2}-\frac{f-2\Box f_{,R}}{f_{R}}+\frac{2H\dot{f}_{,R}}{f_{,R}}+\frac{\ddot{f}_{R}-\dot{\phi}^{2}}{f_{,R}}-\frac{\nabla^{2}}{a^{2}}\right)\\
\times(\Phi+\Psi)=\frac{2\dot{\phi}\delta\dot{\phi}}{f_{,R}}.
 \end{multline}
Finally, by substituting in the right-hand side of this equation the previous result (\ref{intermediatescalar0i}), we find,
\begin{multline}\label{intermediatescalar00+ij}
\ddot{\Phi}+\ddot{\Psi}+\left(H-\frac{2\ddot{\phi}}{\dot{\phi}}+\frac{3\dot{f}_{,R}}{f_{,R}}\right)(\dot{\Phi}+\dot{\Psi})
\\-\left(4\dot{H}-\frac{4\ddot{f}_{,R}}{f_{,R}}+\frac{6\dot{f}_{,R}}{f_{,R}}\frac{\ddot{\phi}}{\dot{\phi}}\right)\Phi
\\+\Bigg[6\dot{H}+6H^{2}+H\left(\frac{4\dot{f}_{,R}}{f_{,R}}-\frac{2\ddot{\phi}}{\dot{\phi}}\right)
-\frac{f-2\Box f_{,R}}{f_{R}}
\\-\frac{\ddot{f}_{,R}}{f_{,R}}+\frac{2\dot{f}_{,R}}{f_{,R}}\frac{\ddot{\phi}}{\dot{\phi}}-\frac{\dot{\phi}^{2}}{f_{,R}}-\frac{\nabla^{2}}{a^{2}}\Bigg](\Phi+\Psi)=0.
\end{multline}

This result is the general dynamical equation of the gravitational potentials $\Phi$ and $\Psi$ in any $f(R,\phi)$ model. We shall apply this equation to the (varying power)-law model in the next section. First, though, we shall find in the next subsection the analogue of this equation for the tensor modes.

\subsection{Tensor modes in $f(R,\phi)$ gravity}\label{TSTensorModes}
In order to study the tensor perturbations, one introduces again a small general perturbation $h_{\mu\nu}$ on the FLRW background metric $\bar{g}_{\mu\nu}$ such that the new metric reads $g_{\mu\nu}=\bar{g}_{\mu\nu}+h_{\mu\nu}$. The non-vanishing perturbed components of the Chritstoffel symbols that will be relevant for us here are then \cite{Weinberg}, $\delta\Gamma_{00}^{0}=-\frac{1}{2}\dot{h}_{00}$, $\delta\Gamma_{0i}^{0}=-\frac{1}{2}\partial_{i}h_{00}$ and $\delta\Gamma^{0}_{ij}=(a^{2}H\delta_{ij}h_{00}-\partial_{(i}h_{j)0}+\frac{1}{2}\dot{h}_{ij})$. Therefore, the $00$-component and the $ij$-components of the above perturbed field equations (\ref{PerturbedFE}), read, respectively,
\begin{multline}\label{Tensor00}
f_{,R}\delta R_{00}-(3\dot{H}+3H^{2}+\partial_{0}^{2})\delta f_{,R}+\frac{1}{2}\delta(f-2\Box f_{,R})
\\-\frac{h_{00}}{2}(f-2\Box f_{,R})-\frac{\dot{h}_{00}}{2}\dot{f}_{,R}=\dot{\phi}\delta\dot{\phi},
\end{multline}
\vspace{-0.5cm}
\begin{multline}\label{Tensorij}
f_{,R}\delta R_{ij}-\frac{h_{ij}}{2}(f-2\Box f_{,R})-\nabla_{i}\nabla_{j}\delta f_{,R}
\\+\delta_{ij}a^{2}\left[(\dot{H}+3H^{2})\delta f_{,R}-\frac{1}{2}\delta(f-2\Box f_{,R})\right]
\\+\left[a^{2}H\delta_{ij}h_{00}-\partial_{(i}h_{j)0}+\frac{1}{2}\dot{h}_{ij}\right]\dot{f}_{,R}=a^{2}\delta_{ij}\dot{\phi}\delta\dot{\phi}.
\end{multline}

In contrast to what we did for the scalar modes, where we were only interested in the gravitational potentials $\Phi$ and $\Psi$, in this subsection, being interested in the gravitational waves, we will decompose the total perturbation $h_{\mu\nu}$ of the metric as follows:
\begin{equation}
\mathrm{d}s^{2}=-(1+2\Phi)\mathrm{d}t^{2}+\left[a^{2}(t)(1-2\Psi)\delta_{ij}+h_{ij}\right]\mathrm{d}x^{i}\mathrm{d}x^{j}.
\end{equation}
By using the transverse gauge, we manage to have a vanishing vector part and a transverse 3-tensor, $\partial_{i}h^{i}_{\,j}=0$, satisfying also the traceless condition, $h_{i}^{\,i}=0$. With this decomposition of the metric, we also have a similar splitting for the linearized perturbation of the Ricci tensor as, $\delta R_{\mu\nu}=\delta R_{\mu\nu}^{h}+\delta R_{\mu\nu}^{S}$. Here, $\delta R_{\mu\nu}^{h}$ is the Ricci tensor constructed from the tensorial components $h_{ij}$ of the metric, while $\delta R_{\mu\nu}^{S}$ is the Ricci tensor built from the scalar components $\Phi$ and $\Psi$. The components of the perturbation $\delta R_{\mu\nu}^{S}$ have already been found previously and are given by (\ref{scalardeltaR0i}), (\ref{scalardeltaR00}) and (\ref{scalardeltaRij}).

With this splitting, it is easy to see that after substitution in the perturbed field equations (\ref{Tensor00}) and (\ref{Tensorij}), one recovers all the terms already found in (\ref{Scalar00}) and (\ref{Scalarij}). However, new additional terms will appear in (\ref{Tensorij}), coming from $\delta R_{ij}^{h}$, as well as the two new terms, $\frac{1}{2}h_{ij}(f-2\Box f_{,R})$ and $\frac{1}{2}\dot{h}_{ij}\dot{f}_{,R}$, that were absent in (\ref{Scalarij}). Therefore, with an analysis similar to the one done for (\ref{Scalar00}) and (\ref{Scalarij}), we conclude that after adding the two lines (\ref{Tensor00}) and (\ref{Tensorij}) the scalar terms $\Phi$ and $\Psi$ could be chosen such that one eliminates all the terms that already appeared in the previous subsection and which were related to the perturbation $\delta f_{,R}$.

Notice that this procedure is reminiscent of what was done in Ref.~\cite{Yang*} for $f(R)$ gravity models, where one introduces the decomposition $g_{\mu\nu}=\bar{g}_{\mu\nu}+h_{\mu\nu}+\bar{g}_{\mu\nu}F$, with $h_{\mu\nu}$ chosen such that it is transverse and traceless~, while the scalar $F$ is chosen such that it cancels out all the contributions of $\delta f_{,R}$ and its derivatives. Here, we had to introduce two functions $\Phi$ and $\Psi$ because in $f(R,\phi)$ models one has, in addition to the Ricci scalar $R$, also the field $\phi$ as a separate degree of freedom.

Thus, after eliminating the gravitational potentials $\Phi$ and $\Psi$ from the sum of (\ref{Tensor00}) and (\ref{Tensorij}) one is left with the following simple set of equations:
\begin{equation}\label{ij+00}
\delta R^{h}_{ij}-h_{ij}\frac{f-2\Box f_{,R}}{2f_{,R}}+\dot{h}_{ij}\frac{\dot{f}_{,R}}{2f_{,R}}=0.
\end{equation}
Here we have made use of the fact that $\delta R^{h}_{00}=0$, as one might easily verify \cite{Weinberg} given the traceless and transverse conditions we imposed on $h_{ij}$. On the other hand, given these traceless and transverse conditions we chose for $h_{ij}$, the perturbation $\delta R^{h}_{ij}$ simplifies to \cite{Weinberg},
\begin{equation}
\delta R^{h}_{ij}=\frac{1}{2}\left(\partial_{0}^{2}-H\partial_{0}+4H^{2}-\frac{\nabla^{2}}{a^{2}}\right)h_{ij}.
\end{equation}
Substituting this inside (\ref{ij+00}), the latter takes the following form:
\begin{equation}\label{ij+00final}
\left[\partial_{0}^{2}\!-\!\left(H-\frac{\dot{f}_{,R}}{f_{,R}}\right)\!\partial_{0}+4H^{2}-\frac{f-2\Box f_{,R}}{f_{,R}}-\frac{\nabla^{2}}{a^{2}}\right]\!h_{ij}=0.
\end{equation}

This second-order differential equation in time $t$ is the equation governing the dynamics of the transverse part $h_{ij}$ of the background metric perturbation in a general $f(R,\phi)$ model of gravity.

\section{Application to the (varying power)-law model}\label{sec:3}
Let us now use all the results found in the previous section for general $f(R,\phi)$-gravity to deduce the consequences of the (varying power)-law model on the density perturbations. Let us first recall the main equations of the model. By setting $c=\hbar=1$ and introducing the mass scale $\mu$, the gravitational action of the (varying power)-law model reads~\cite{Hammad2},
\begin{align}\label{Action}
S=\frac{M_{P}^{2}}{16\pi}\int\mathrm{d}^{4}x\sqrt{-g}\left[R-(\partial\phi)^2
-m^{2}\phi^{2}-\frac{\mu^{2}}{2}\left(\frac{R}{\mu^{2}}\right)^{\phi}\right],
\end{align}
where $M_{P}$ is the Planck mass and $m$, whose order of magnitude is not constrained by the model, is the scalar field's mass.

Notice that here we brought two modifications inside the action with respect to the one used in Refs.~\cite{Hammad2,Hammad3}. The first modification is that we have chosen the mass scale $\mu$ in the denominator of $(R/\mu^2)^\phi$ instead of the higher Planck scale $M_{P}^2$ chosen in Refs.~\cite{Hammad2,Hammad3}. The second modification is that we have divided by 2 the factor $\mu^2$ in front of the term $(R/\mu^2)^\phi$. As we will see shortly, the second modification is dictated by the first modification. The reason for the first modification will be explained later on when we examine the power spectra in subsection \ref{VPLPowers}. Note, however, that the analysis done in Ref.~\cite{Hammad3} concerning the effective mass of the scalar field that rises from the radiative corrections in the model remains valid since the analysis done there did not depend at all on the ratio $R/\mu^{2}$.

From the action (\ref{Action}) it is clear that the potential of the scalar field $\phi$ depends also on the Ricci scalar and reads,
\begin{align}\label{Potential}
V(R,\phi)=\frac{m^{2}}{2}\phi^{2}+\frac{\mu^{2}}{4}\left(\frac{R}{\mu^{2}}\right)^{\phi}.
\end{align}
First, the full modified Friedmann equation one finds by writing the field equations in the flat FLRW metric is \cite{Hammad2},
\begin{multline}\label{Friedmann}
\frac{\phi}{2}\left(\frac{R}{\mu^{2}}\right)^{\phi-1}\dot{H}+\left[1+\frac{\phi}{2}\left(\frac{R}{\mu^{2}}\right)^{\phi-1}\right]H^{2}=
\\\frac{m^{2}\phi^{2}}{6}+\frac{\mu^{2}}{12}\left(\frac{R}{\mu^{2}}\right)^{\phi}+\frac{\dot{\phi}^{2}}{6}
+\frac{\mu^{2}H}{2}\frac{\mathrm{d}}{\mathrm{d}t}\left[\frac{\phi}{R}\left(\frac{R}{\mu^{2}}\right)^{\phi}\right].
\end{multline}
Here, $H=\dot{a}/a$ is the Hubble parameter and an overdot denotes, as usual, a time derivative. From this equation we see the reason why we divided here by 2 the mass squared $\mu^2$ in front of $(R/\mu^{2})^{\phi}$ in the action (\ref{Action}). Indeed, setting $R=\mu^2$ in (\ref{Friedmann}) and $\dot{\phi}=\dot{H}=0$ gives $H^2=\mu^2/12$. This is consistent with the fact that $R=12H^2$ when $\dot{H}=0$. This would have been otherwise if we had not divided by 2 the factor $\mu^{2}$ in the action.

Next, the Klein-Gordon equation one obtains from the field equations is \cite{Hammad2},
\begin{align}\label{KG}
\ddot{\phi}+3H\dot{\phi}+\frac{\partial V(R,\phi)}{\partial{\phi}}=0.
\end{align}
In order to substitute for $\dot{\phi}$ arising in the Friedmann equation (\ref{Friedmann}) the Klein-Gordon equation needs to be solved for $\dot{\phi}$ in terms of $\phi$ and $R$. To be able to do that, however, the simplest way is to discard the second derivative $\ddot{\phi}$ from (\ref{KG}) and keep only the other two terms. This is possible provided that either $|\ddot{\phi}|$ can be neglected in front of the Hubble friction term $3H|\dot{\phi}|$ or that the ratio $|\ddot{\phi}|/3H|\dot{\phi}|$ stays nearly constant so that one could turn (\ref{KG}) into a first-order differential equation in $\phi$. As we shall see below, this second condition is satisfied during inflation.

Note, however, that in contrast to usual single-field inflationary models, where one requires the condition $|\ddot{\phi}|\ll3H|\dot{\phi}|$ to sustain inflation, our condition to have $|\ddot{\phi}|$ small enough to be neglected in front of $3H|\dot{\phi}|$ or just be proportional to the latter is not required by inflation itself but represents only a possible simplifying assumption that will turn out to be automatically satisfied by the model during inflation.

Instead of relying on such assumptions, in what follows we are going to keep, as a first step, all the terms in (\ref{KG}) and write the latter as,
\begin{align}\label{ApproximateKG}
\dot{\phi}=-\frac{1}{3H}\frac{\partial V}{\partial\phi}-\frac{\ddot\phi}{3H}.
\end{align}

We now set, for convenience, $R/\mu^{2}=\rho$. Then, after neglecting the mass term $m^2\phi^2$ in the potential (\ref{Potential}) and using the approximation $R\approx12H^{2}$, extracted from the geometric identity $R=6\dot{H}+12H^{2}$ for an FLRW Universe, as well as the approximation $\dot{R}=6\ddot{H}+24H\dot{H}\approx24H\dot{H}$, we find, at the first order in $\dot H/H^2$, the following approximation for $\dot{\phi}$:
\begin{align}\label{UsefulKG}
\dot{\phi}\simeq-\left(H+\frac{\dot{H}}{2H}\right)\rho^{\phi-1}\ln\rho-\frac{\ddot\phi}{3H}.
\end{align}

Substituting this approximation for $\dot{\phi}$ as well as the above approximations for $R$ and $\dot{R}$, the modified Friedmann equation (\ref{Friedmann}) becomes, at the first order in $\dot H/H^2$ and $\ddot\phi/H^2$, as follows:
\begin{multline}\label{ApproximateFriedmann}
2+(\phi-2)\rho^{\phi-1}-\left(\frac{1}{3}\ln^{2}\rho-\ln\rho-\phi\ln^{2}\rho\right)\rho^{2\phi-2}=\\
\frac{\dot{H}}{2H^2}\bigg[\left(\frac{2-3\phi}{3}\ln^{2}\rho-\ln\rho\right)\rho^{2\phi-2}
+(2+4\phi^{2}-6\phi)\rho^{\phi-1}\bigg]\\+\frac{\ddot\phi}{3H^2}\left(\frac{2-3\phi}{3}\ln\rho-1\right)\rho^{\phi-1}.
\end{multline}

Now we need to find the corresponding value of the field $\phi$ for each value of the ratio $\rho$; that is, for each value of the Hubble parameter $H$ during inflation. This is possible only if we neglect the ratios $\dot{H}/H^2$ and $\ddot{\phi}/H^2$ in the above equation and then solve the resulting equation for different energy scales. Just as in usual single-field inflationary models, this approximation is justified by the fact that for inflation to last long enough the variation of the Hubble parameter, as well as the variation of $\dot\phi$, should be insignificant relative to $H^2$ during inflation. Furthermore, as it will turn out, this approximation will be justified {\textit{a posteriori}} as revealed by the third and fifth columns of Table \ref{tab:Table} below, making our mathematical procedure fully consistent. The equation that one obtains is thus,
\begin{equation}\label{RhoEquation}
2+(\phi-2)\rho^{\phi-1}-\left(\frac{1}{3}\ln^{2}\rho-\ln\rho-\phi\ln^{2}\rho\right)\rho^{2\phi-2}\simeq0.
\end{equation}

Note that, contrary to familiar single-field inflationary models, the Hubble parameter squared $H^2$ in this model is not simply given by the scalar field's potential (\ref{Potential}), but rather by a more complicated effective potential $H^2=\cal{V}(\phi)$ that cannot be extracted analytically as we see from the highly non-linear equation (\ref{RhoEquation}).

Note also that for $\dot{H}=0$ and $\rho=1$, \textit{i.e.}, for $R=\mu^{2}$, all the logarithm terms in this equation vanish and the equation gives $\phi=0$. That is, at the inflation scale $\mu^2$ the value of the scalar field vanishes and starts to increase with the expansion of the Universe from then on during inflation.

Since equation (\ref{RhoEquation}) allows us only to find $H$ and $\phi$, we need to differentiate the equation once more with respect to time $t$ in order to find the approximate values for the ratio $\dot{H}/H^{2}$ that we neglected in (\ref{ApproximateFriedmann}). Differentiating (\ref{RhoEquation}) and then substituting in the resulting equation $\dot{\phi}$ from (\ref{UsefulKG}), we find, after setting $\ln\rho=\ell$ for convenience,
\begin{multline}\label{HDotOverH2}
\frac{\dot{H}}{H^2}\simeq\left[\rho^{\phi-1}+\left(\ell^2+\phi\ell^3-\frac{\ell^3}{3}\right)\rho^{2\phi-2}-2\ell\right]\times\\
\Bigg[\frac{4(1-\phi)}{\ell}\rho^{1-\phi}+\ell+\left(\frac{\ell^3}{6}-\frac{\phi\ell^3}{2}-\frac{\ell^2}{2}\right)\rho^{2\phi-2}
\\-\left(\frac{17}{6}-\frac{2\ell}{3}-6\phi+\frac{8\phi\ell}{3}-2\phi^2\ell
-\frac{2}{\ell}\right)\rho^{\phi-1}\Bigg]^{-1}.\!\!\!\!
\end{multline}
As we see from equations (\ref{intermediatescalar00+ij}) and (\ref{ij+00final}) of Section~\ref{sec:2}, we will also need hereafter the ratio $\ddot{\phi}/H^2$. Differentiating (\ref{UsefulKG}) once with respect to time, we find,
\begin{multline}\label{PhiDDotOverH2}
\frac{\ddot{\phi}}{H^2}\simeq\rho^{\phi-1}\Bigg[\left(\ell-2-2\phi\ell+\frac{\ell^3}{2}\rho^{\phi-1}\right)\frac{\dot{H}}{H^2}\\
-\left(1-\frac{3\ell}{2}+\phi\ell\right)\frac{\dot{H}^2}{H^4}-\frac{\ddot{H}\ell}{2H^3}+\ell^3\rho^{\phi-1}\Bigg].
\end{multline}

Now, the numerical values of $\rho$ and $\phi$, as well as the ratios $\dot{\phi}/H$, $\dot{H}/H^2$ and $\ddot{\phi}/H^2$ will all be needed in order to compute the quantities that would yield the density perturbations to be compared with observations. For that purpose, we have tabulated the numerical values that correspond to the various energy scales immediately below the scale $\mu^{2}$ using equations (\ref{RhoEquation}), (\ref{UsefulKG}), (\ref{HDotOverH2}) and (\ref{PhiDDotOverH2}), respectively.

One criterion used in usual single-field inflationary models, and that remains valid in the present model, is to satisfy the so-called first two slow-roll conditions. These conditions guarantee that the rapid expansion of the early Universe is maintained sufficiently long for the Universe to reach the size required to solve the horizon and flatness problems \cite{Riotto}. Indeed, these conditions are independent of the particular model one chooses as they represent conditions on the kinematics of the early Universe. The first condition is the one we already saw above, which is that the variation of the Hubble parameter be negligible, while the second condition is that the variation of the field $\phi$ be negligible in front of $H$. That is,
\begin{equation}\label{Epsilon-Eta}
\epsilon\equiv\frac{|\dot{H}|}{H^{2}}\ll1,\qquad\qquad\eta\equiv\frac{|\dot{\phi}|}{H}\ll1.
\end{equation}

In usual inflationary models, where the Hubble parameter is given by $H^{2}\sim V(\phi)$, the analogue of the $\eta$-condition is $\dot{\phi}^2/H^2\ll1$, which, thanks to (\ref{ApproximateKG}), is just equivalent to the $\epsilon$-condition in (\ref{Epsilon-Eta}). As we saw above, however, in the (varying power)-law model the Hubble parameter is not simply given by $H^{2}\sim V(\phi)$ and, therefore, no condition should \textit{a priori} be imposed on $\dot{\phi}$. As we shall see below, though, our $\eta$-condition will be automatically satisfied throughout all the duration of inflation.

The different values we find for different energy scales using a numerical computation are tabulated below.

\renewcommand{\arraystretch}{1.5}
\begin{table*}
\begin{center}
\begin{tabular}{||c c | c c c c||c ||c c| c c c c ||}
  \cline{1-6}\cline{8-13}
      $\rho$     &   $\phi$   & $\dot{H}/H^2$ & $\dot{\phi}/H$ & $\ddot{\phi}/H^2$ & $\ddot{\phi}/H\dot{\phi}$ &$ $ & $\rho$     &   $\phi$   & $\dot{H}/H^2$ & $\dot{\phi}/H$ & $\ddot{\phi}/H^2$ & $\ddot{\phi}/H\dot{\phi}$\\
  \cline{1-6}\cline{8-13}
  \cline{1-6}\cline{8-13}
  $10^{-10^{-8}}$ & $7\times10^{-8}$ & $-4\times10^{-9}$ & $2\times10^{-8}$  & $8\times10^{-9}$ & 0.33  & &   $10^{-0.0060}$ &  0.040396  &  -0.002458    &  0.013983      &   0.005014        &    0.36  \\
  $10^{-10^{-7}}$ & $7\times10^{-7}$ & $-4\times10^{-8}$ & $2\times10^{-7}$  & $8\times10^{-8}$ & 0.33  & &   $10^{-0.0070}$ &  0.046932  &  -0.002899    &  0.016344      &   0.005930        &    0.36  \\
  $10^{-10^{-6}}$ & $7\times10^{-6}$ & $-4\times10^{-7}$ & $2\times10^{-6}$  & $8\times10^{-7}$ & 0.33  & &   $10^{-0.0080}$ &  0.053413  &  -0.003348    &  0.018713      &   0.006870        &    0.37  \\
  $10^{-10^{-5}}$ & 0.000069  &  -0.000004    &  0.000023      &   0.000008        &    0.33            & &   $10^{-0.0090}$ &  0.059840  &  -0.003806    &  0.021091      &   0.007832        &    0.37  \\
  $10^{-0.0001}$ &  0.000690  &  -0.000038    &  0.000230      &   0.000079        &    0.33            & &   $10^{-0.0100}$ &  0.066214  &  -0.004273    &  0.023476      &   0.008819        &    0.38  \\
  $10^{-0.0002}$ &  0.001380  &  -0.000077    &  0.000460      &   0.000154        &    0.33            & &   $10^{-0.0110}$ &  0.072537  &  -0.004749    &  0.025869      &   0.009828        &    0.38  \\
  $10^{-0.0003}$ &  0.002070  &  -0.000116    &  0.000691      &   0.000231        &    0.33            & &   $10^{-0.0120}$ &  0.078807  &  -0.005234    &  0.028269      &   0.010862        &    0.38  \\
  $10^{-0.0004}$ &  0.002758  &  -0.000154    &  0.000922      &   0.000309        &    0.34            & &   $10^{-0.0130}$ &  0.085027  &  -0.005227    &  0.030677      &   0.011919        &    0.39  \\
  $10^{-0.0005}$ &  0.003446  &  -0.000193    &  0.001153      &   0.000387        &    0.34            & &   $10^{-0.0140}$ &  0.091196  &  -0.006230    &  0.033091      &   0.033001        &    0.39  \\
  $10^{-0.0006}$ &  0.004134  &  -0.000232    &  0.001383      &   0.000465        &    0.34            & &   $10^{-0.0150}$ &  0.097316  &  -0.006743    &  0.035512      &   0.014106        &    0.40  \\
  $10^{-0.0007}$ &  0.004821  &  -0.000271    &  0.001614      &   0.000543        &    0.34            & &   $10^{-0.0160}$ &  0.103387  &  -0.007264    &  0.037940      &   0.015235        &    0.40  \\
  $10^{-0.0008}$ &  0.005507  &  -0.000310    &  0.001845      &   0.000621        &    0.34            & &   $10^{-0.0170}$ &  0.109409  &  -0.007794    &  0.040375      &   0.016389        &    0.41  \\
  $10^{-0.0009}$ &  0.006193  &  -0.000349    &  0.002076      &   0.000700        &    0.34            & &   $10^{-0.0180}$ &  0.115383  &  -0.008334    &  0.042815      &   0.017567        &    0.41  \\
  $10^{-0.0010}$ &  0.006878  &  -0.000388    &  0.002307      &   0.000779        &    0.34            & &   $10^{-0.0190}$ &  0.121310  &  -0.008883    &  0.045262      &   0.018769        &    0.41  \\
  $10^{-0.0020}$ &  0.013697  &  -0.000785    &  0.004624      &   0.001580        &    0.34            & &   $10^{-0.0200}$ &  0.127190  &  -0.009442    &  0.047714      &   0.019996        &    0.42  \\
  $10^{-0.0030}$ &  0.020457  &  -0.001190    &  0.006951      &   0.000079        &    0.35            & &   $10^{-0.0300}$ &  0.183521  &  -0.015556    &  0.072517      &   0.033640        &    0.46  \\
  $10^{-0.0040}$ &  0.027160  &  -0.001604    &  0.009286      &   0.003251        &    0.35            & &   $10^{-0.0400}$ &  0.235664  &  -0.022677    &  0.097700      &   0.049857        &    0.51  \\
  $10^{-0.0050}$ &  0.033806  &  -0.002027    &  0.011630      &   0.004121        &    0.35            & &   $10^{-0.0500}$ &  0.283988  &  -0.030880    &  0.123092      &   0.068763        &    0.56  \\
  \cline{1-6}\cline{8-13}
\end{tabular}
\caption{The numerical values of the various parameters at different energy scales.}\label{tab:Table}
\end{center}
\end{table*}
From this table we clearly see that in order to satisfy the two slow-roll conditions enumerated above, inflation in this model should start at an energy scale $\mu$ and end at another energy scale not very far off. The end of inflation will occur at that energy scale for which either both or one of the slow-roll parameters in (\ref{Epsilon-Eta}) ceases to be small and negligible. It is clear from this table that both slow-roll conditions in (\ref{Epsilon-Eta}) are satisfied up to the scale $\rho\sim10^{-0.05}$. Around the energy scale $\rho\sim10^{-0.05}$, the second slow-roll condition on $\eta$ starts to be violated. Therefore, in this model inflation should end at this energy scale.

In fact, in contrast to what is done in usual scalar field inflationary models, we will not be able to compute here the number of $e$-folds $N$ to deduce the required final value of the scalar field at the end of inflation. This is because the number of $e$-folds $N$ would have to be found from the following integral:
\begin{equation}\label{Efolds1}
N=\int_{i}^{f}H\mathrm{d}t=-3\int_{\phi_{i}}^{\phi_{f}}\frac{\mathcal{V}(\phi)}{V_{,\phi}(\phi)}\mathrm{d}\phi.
\end{equation}
As we saw above, unfortunately, the analytic evaluation of this integral is rendered impossible because no analytic expression for the effective potential $\mathcal{V}(\phi)$ could be extracted from (\ref{RhoEquation}).

It is still possible, however, to estimate the value of the integral (\ref{Efolds1}) by making some approximations based on the results we found above. Indeed, using the result (\ref{HDotOverH2}) with $\rho\sim1$ and $\phi\ll1$, we can use the following approximation: $\epsilon\simeq-\ln\rho/6$. On the other hand, since $12H^2\simeq R(1-\frac{1}{2}\epsilon)$, we also have $\mathrm{d}\ln H\simeq\frac{1}{2}\mathrm{d}\ln\rho-\frac{1}{4}\mathrm{d}\epsilon$. Therefore, the first integral on the right-hand side in (\ref{Efolds1}), which can be written as $\int_{i}^{f}H\mathrm{d}t=-\int_{i}^{f}\mathrm{d}\ln H/\epsilon$, can also be performed as follows:
\begin{equation}\label{Efolds2}
N\simeq\int_{i}^{f}\left(3\frac{\mathrm{d}\ln\rho}{\ln\rho}+\frac{\mathrm{d}\epsilon}{4\epsilon}\right)
=\frac{13}{4}\ln\left(\frac{\ln\rho_{f}}{\ln\rho_{i}}\right).
\end{equation}

With this formula, we can easily investigate the possible initial and final values of $\rho$ that would lead to the necessary amount of $e$-folds expected from any inflationary model. The required number of $e$-folds to agree with observation is generally set to be at least $N\simeq50$ \cite{Planck2015}. From the table above, we easily verify that the closer to the $\mu$-scale one allows inflation to end, the closer to that scale one needs inflation to also start. For example, if one assumes inflation to have ended at the energy scale of $\rho\simeq10^{-0.05}$, then one should allow inflation to start at the energy scale of $\rho\simeq10^{-10^{-8}}$, in which case one finds, $N\simeq50$. If, on the other hand, one assumes inflation to have ended at the energy scale of $\rho\simeq10^{-0.01}$, then one finds for the same starting energy scale, $N\simeq45$. To have a bigger number of $e$-folds one needs to push back the starting of inflation to scales which are even closer to the $\mu$-scale, if not exactly equal to $\mu$.

With these numerical results, we shall investigate in the next two subsections the density perturbations as implied by the model in detail.

\subsection{Scalar modes in the (varying power)-law model}\label{VPLScalarModes}
Let us now adapt all the equations found in Section \ref{sec:2} to the (varying power)-law model. First of all, since in this model $f(R,\phi)=R-\frac{1}{2}\mu^2(R/\mu^{2})^{\phi}$, we have $f_{,R}=1-\frac{1}{2}\phi(R/\mu^{2})^{\phi-1}$. On the other hand, since inflation happens for $\mu^{2}\sim R=6\dot{H}+12H^{2}$, as we saw below equation (\ref{RhoEquation}), and for which we found $\phi\ll1$ as it appears from Table \ref{tab:Table} of the previous subsection, we may adopt the following approximations: $f\sim R-\frac{1}{2}\mu^{2}\sim3\dot{H}+6H^{2}$, $f_{,R}\sim 1$, $\dot{f}_{,R}\sim-\frac{1}{2}\dot{\phi}$, $\ddot{f}_{,R}\sim-\frac{1}{2}\ddot{\phi}$ and $\Box f_{,R}\sim-\frac{1}{2}\Box\phi$. Then, by substituting these inside the differential equation (\ref{intermediatescalar00+ij}), the latter takes, in Fourier space, the following form:
\begin{multline}\label{beforefinalscalar00+ij}
\!\!\ddot{\Phi}_{\mathbf{k}}+\ddot{\Psi}_{\mathbf{k}}+\left(H-\frac{2\ddot{\phi}}{\dot{\phi}}-\frac{3\dot{\phi}}{2}\right)(\dot{\Phi}_{\mathbf{k}}+\dot{\Psi}_{\mathbf{k}})-\left(\dot{H}-\frac{\ddot{\phi}}{4}\right)\delta\phi_{\mathbf{k}}\\
+\left[\dot{H}+H\left(\dot{\phi}-\frac{2\ddot{\phi}}{\dot{\phi}}\right)-\dot{\phi}^{2}+\frac{k^{2}}{a^{2}}\right](\Phi_{\mathbf{k}}+\Psi_{\mathbf{k}})=0.
\end{multline}
Here, we have used the fact that $\Psi-\Phi=\delta f_{,R}/f_{,R}\sim-\frac{1}{2}\delta\phi$, as it follows from the approximations we made above, in order to recast the differential equation in terms of the sum $\Phi+\Psi$. On the other hand, equation (\ref{simplescalar0i}) becomes, within our approximations for $f_{,R}$ and $\dot{f}_{,R}$, as follows:
\begin{equation}\label{Approxsimplescalar0i}
\dot{\Phi}+\dot{\Psi}+\left(H-\frac{\dot{\phi}}{4}\right)(\Phi+\Psi)=\frac{11}{8}\dot{\phi}\delta\phi.
\end{equation}
Therefore, by combining equations (\ref{beforefinalscalar00+ij}) and (\ref{Approxsimplescalar0i}), and setting $\Psi+\Phi=\Theta$, we obtain,
\begin{multline}\label{finalscalar00+ij}
\ddot{\Theta}_{\mathbf{k}}+H\left(1-\frac{20\ddot{\phi}}{11H\dot{\phi}}-\frac{8\dot{H}}{11H\dot{\phi}}-\frac{3\dot{\phi}}{2H}\right)\dot{\Theta}_{\mathbf{k}}\\
-H^{2}\Bigg(\frac{20\ddot\phi}{11H\dot\phi}+\frac{8\dot H}{11H\dot\phi}+\frac{21\ddot\phi}{22H^2}+\frac{\dot\phi^2}{H^2}-\frac{\dot\phi}{H}
\\-\frac{13\dot H}{11H^2}-\frac{k^2}{a^2H^2}\Bigg)\Theta_{\mathbf{k}}=0.
\end{multline}
Let us distinguish the two regimes of sub-horizon and super-horizon scales for which one has, respectively, $k\gg aH$ and $k\ll aH$. Here, we are interested in the latter. In this case, the last term in the second set of  parentheses of (\ref{finalscalar00+ij}) can be neglected. Furthermore, according to the results in Table \ref{tab:Table}, all the other terms in both sets of parentheses can also be safely neglected except for the two fractions $\ddot\phi/H\dot\phi$ and $\dot H/H\dot\phi$. However, Table \ref{tab:Table} shows that these two fractions remain approximately constant during all the period we take for inflation. Therefore, we learn that, to a good approximation, Eq. (\ref{finalscalar00+ij}) is of the form,
\begin{equation}
\ddot\Theta_{\mathbf{k}}+H(1-A)\dot{\Theta}_{\mathbf{k}}-AH^{2}\Theta_{\mathbf{k}}=0,
\end{equation}
where, $A$ is a positive constant because $|\ddot\phi/H\dot\phi|>|\dot H/H\dot\phi|$. The solution to this last equation is of the form, $\Theta_{\mathbf{k}}=c_{1}(k)e^{-Ht}+c_{2}(k)e^{AHt}$. Keeping the converging solution, we set the second integration constant to vanish, $c_{2}(k)=0$, while we keep $c_{1}(k)$ arbitrary. Therefore, we learn that for super-horizon scales the sum $\Psi+\Phi$ is given by,
\begin{equation}
\Psi+\Phi\sim\frac{c_{1}(k)}{a}=\frac{c_{1}(k)H}{k}.
\end{equation}
Here, we have used the fact that at Hubble crossing, $k=aH$. We deduce from this that for the quasi-de Sitter regime, $\dot{\Psi}_{\mathbf{k}}+\dot{\Phi}_{\mathbf{k}}\simeq-\epsilon H(\Psi_{\mathbf{k}}+\Phi_{\mathbf{k}})$. Using this result in (\ref{Approxsimplescalar0i}), we have finally that,
\begin{equation}\label{Psi+Phi}
\Psi_{\mathbf{k}}+\Phi_{\mathbf{k}}\simeq\frac{13\eta}{8}\delta\phi_{\mathbf{k}}.
\end{equation}
What we are interested in actually is the gauge invariant curvature perturbation $\mathcal{R}$ that combines the metric and matter field perturbations, $\mathcal{R}=\Psi+H\delta\phi/\dot{\phi}$ (see \textit{e.g.}, Refs.~\cite{Mukhanov,Riotto}.) This gauge-invariant quantity gives the curvature perturbation in co-moving coordinates because it is constructed so that it allows one to find the curvature perturbation on hypersurface slices such that $\delta\phi=0$. Now, since we already found that $\Phi=\Psi+\frac{1}{2}\delta\phi$, we deduce from (\ref{Psi+Phi}) that the gauge-invariant curvature perturbation is simply $\mathcal{R}\simeq H\delta\phi/\dot{\phi}$. This is in agreement with what single-field inflationary models predict within general relativity.

All that remains then is to find the modes $\delta\phi_{\mathbf{k}}$ of the scalar perturbation. For that, let us go back to the Klein-Gordon equation (\ref{KG}) and find the equation obeyed by the perturbation $\delta\phi$. Perturbing that equation gives, in Fourier space,
\begin{multline}\label{PerturbedKG}
\delta\ddot{\phi}_{\mathbf{k}}+3H\delta\dot{\phi}_{\mathbf{k}}+\left(V_{,\phi\phi}+\frac{k^{2}}{a^{2}}\right)\delta\phi_{\mathbf{k}}
\\=-V_{,\phi R}\delta R_{\mathbf{k}}-2\Phi_{\mathbf{k}}V_{,\phi}+(\dot{\Phi}_{\mathbf{k}}+3\dot{\Psi}_{\mathbf{k}})\dot{\phi}.
\end{multline}
We should eliminate from this equation the perturbation $\delta R_{\mathbf{k}}$ of the Ricci scalar on the right-hand side by expressing it in terms of the perturbation $\delta\phi_{\mathbf{k}}$. Let us write down the expression of the perturbed Ricci scalar, $\delta R$. This could be done either by using the perturbed metric (\ref{NewtonianGauge}) or by just combining (\ref{scalardeltaR00}) and (\ref{scalardeltaRij}). We find the following result in Fourier space:
\begin{multline}\label{DeltaR}
\delta R_{\mathbf{k}}=-6\ddot{\Psi}_{\mathbf{k}}-H(21\dot{\Psi}_{\mathbf{k}}+9\dot{\Phi}_{\mathbf{k}})
-(6\dot{H}+18H^2)(\Psi_{\mathbf{k}}+\Phi_{\mathbf{k}})
\\-\frac{k^{2}}{a^{2}}(4\Psi_{\mathbf{k}}-2\Phi_{\mathbf{k}}).
\end{multline}
By using (\ref{Psi+Phi}) and the fact that $\Psi_{\mathbf{k}}-\Phi_{\mathbf{k}}=-\frac{1}{2}\delta\phi_{\mathbf{k}}$, identity (\ref{DeltaR}) may be approximated to,
\begin{multline}\label{DeltaRApproximated}
\delta R_{\mathbf{k}}=\left(\frac{3}{2}-\frac{39\eta}{8}\right)\delta\ddot{\phi}_{\mathbf{k}}
+3H\left(1-\frac{65\eta}{8}\right)\delta\dot{\phi}_{\mathbf{k}}
\\+\left[\frac{k^{2}}{a^{2}}\left(\frac{3}{2}-\frac{13\eta}{8}\right)-\frac{117\eta}{4}H^2\right]\delta\phi_{\mathbf{k}}.
\end{multline}
Substituting this in the right-hand side of (\ref{PerturbedKG}) after deducing from (\ref{Potential}) that $V_{,\phi R}\sim1/4$ and after using our definition of the slow-roll parameters (\ref{Epsilon-Eta}), we find, up to the first-order in the slow-roll parameters the following differential equation:
\begin{multline}\label{PerturbedKG2}
\delta\ddot{\phi}_{\mathbf{k}}+\frac{30H}{11}\left(1-\frac{501\eta}{440}\right)\delta\dot{\phi}_{\mathbf{k}}
\\+\left[\frac{k^2}{a^2}\left(1+\frac{13\eta}{22}\right)-\frac{141\eta}{16}H^2\right]\delta\phi=0.
\end{multline}
This is the dynamical equation for the scalar perturbation modes. This equation can most easily be solved by switching to conformal time $\tau$, defined by $\mathrm{d}\tau=\mathrm{d}t/a(t)$. For then the equation becomes, after introducing the new variable $\delta\sigma_\mathbf{k}=a\delta\phi_\mathbf{k}$, as follows:
\begin{multline}\label{SigmaEquation}
\delta\sigma''_{\mathbf{k}}+\frac{3}{11\tau}\left(1+\epsilon-\frac{501\eta}{44}\right)\delta\sigma'_{\mathbf{k}}
\\+\left[k^2\left(1+\frac{13\eta}{22}\right)-\frac{19}{11\tau^2}\left(1+\frac{27\epsilon}{19}+\frac{11049\eta}{3344}\right)\right]
\delta\sigma_{\mathbf{k}}=0.\!\!\!\!\!\!
\end{multline}
Here, a prime denotes a derivative with respect to conformal time $\tau$. To obtain this last equation we have assumed a quasi-de Sitter regime for inflation in which case one can approximate the Hubble parameter $H$ in terms of the scale factor $a$ and the conformal time $\tau$ as follows: $H\simeq -(1+\epsilon)/a\tau$\footnote{Note that, as pointed out by the anonymous referee, according to Table I, one does not necessarily have $\dot\epsilon\sim\epsilon^2H$ which allows one to neglect the variation of $\epsilon$ and write $H\simeq-(1+\epsilon)/a\tau$. However, the usual integration that leads to such an approximation, namely, $-1/aH=\int(1-\epsilon)d\tau\simeq(1-\epsilon)\tau-\int\frac{\dot\epsilon}{H}d\tau$, shows that even if $\dot\epsilon$ is not second order in $\epsilon$, the ratio $\frac{\dot\epsilon}{H}$ can, for our purposes, be safely neglected in such an integral.}. Now the generic solution to equation (\ref{SigmaEquation}) is of the form (see \textit{e.g.}, \cite{Bowman}),
\begin{equation}
\delta\sigma_{\mathbf{k}}=(-\tau)^{\alpha_{s}}\left[c_{1}(k)H^{(1)}_{\nu_{s}}(-\beta k\tau)+c_{2}(k)H^{(2)}_{\nu_{s}}(-\beta k\tau)\right],
\end{equation}
where, $c_{1}(k)$ and $c_{2}(k)$ are new integration constants, and $H^{(1)}_{\nu_s}$ and $H^{(2)}_{\nu_s}$ are the Hankel's functions of the first and second kind, respectively. The parameters $\alpha_{s}$, $\nu_{s}$ and $\beta$ are given by,
\begin{align}\label{ScalarAlphaNu}
\alpha_{s}&=\frac{4}{11}-\frac{3\epsilon}{22}+\frac{1501\eta}{968},\quad\nu^2_{s}=\frac{225}{121}+\frac{285\epsilon}{121}+\frac{145555\eta}{21296},
\nonumber\\\beta&=1+\frac{13\eta}{22}.
\end{align}
Since for $x\gg1$ the Hankel's functions satisfy $H^{(1)}_{\nu_s}(x)\sim x^{-\frac{1}{2}}e^{ix}$ and $H^{(2)}_{\nu_s}(x)\sim x^{-\frac{1}{2}}e^{-ix}$ \cite{Bowman}, we shall set the second arbitrary integration constant $c_{2}(k)$ equal to zero while we choose $c_1(k)=k^{\alpha_s-\frac{1}{2}}$ for the first in order to recover the right dimensions for the quantum Fourier modes. On the other hand, since for $x\ll1$ we know that $H_{\nu_s}^{(1)}(x)\sim x^{-\nu_s}$ \cite{Bowman}, we deduce, up to an unimportant multiplicative constant, the following result for super-horizon scales for which $k\ll aH$ ($-k\tau\ll1$):
\begin{equation}
\delta\sigma_{\mathbf{k}}\sim\frac{1}{\sqrt{2k}}(-k\tau)^{\alpha_{s}-\nu_{s}}.
\end{equation}
Therefore, the sought-after solution on super-horizon scales $\delta\phi_{\mathbf{k}}$ reads,
\begin{equation}\label{ScalarModes}
\delta\phi_{\mathbf{k}}\sim\frac{H}{\sqrt{2k^3}}\left(\frac{k}{aH}\right)^{1+\alpha_{s}-\nu_{s}}.
\end{equation}
Before using this solution to compute the power spectrum of the scalar perturbations, we first do a similar analysis in the next subsection and extract the dynamical equations of the tensor modes as well as their final expression.

\subsection{Tensor modes in the (varying power)-law model}\label{VPLTensorModes}
We shall now apply the results obtained in subsection \ref{TSTensorModes} to the case where $f(R,\phi)=R-\frac{1}{2}\mu^{2}(R/\mu^{2})^{\phi}$. As we did in the first paragraph of subsection \ref{VPLScalarModes}, we can use again here the following approximations: $f\sim R-\frac{1}{2}\mu^{2}\sim3\dot{H}+6H^{2}$, $f_{,R}\sim 1$, $\Box f_{,R}\sim-\frac{1}{2}\Box\phi$ and $\dot{f}_{,R}\sim-\frac{1}{2}\dot{\phi}$. Then, by substituting for $\dot{\phi}$ its expression from the approximate Klein-Gordon equation (\ref{ApproximateKG}), the differential equation (\ref{ij+00final}) takes, in Fourier space, the following more explicit form:
\begin{equation}\label{IntermediateWaveEquation}
\left[\partial_{0}^{2}-\left(H+\frac{\dot{\phi}}{2}\right)\partial_{0}-2H^{2}-3\dot{H}
-\Box\phi+\frac{k^{2}}{a^{2}}\right]h_{ij\mathbf{k}}=0.
\end{equation}
After absorbing the scale factor from the tensor $h_{ij}$ by defining the physical tensor mode $\mathcal{H}_{ij}$ \cite{Riotto} such that $h_{ij}=a^{2}\mathcal{H}_{ij}$, and using the fact that $\Box\phi=V_{,\phi}$ as it follows from the Klein-Gordon equation (\ref{KG}), the above equation reduces to the following approximation for the wave equation of the transverse tensor modes:
\begin{equation}\label{WaveEquation}
\ddot{\mathcal{H}}_{ij\mathbf{k}}+3H\left(1-\frac{\eta}{6}\right)\dot{\mathcal{H}}_{ij\mathbf{k}}+\left[\frac{k^{2}}{a^{2}}
+H^{2}\left(2\eta+\epsilon\right)\right]\mathcal{H}_{ij\mathbf{k}}=0.
\end{equation}
Here, we have used again our definition (\ref{Epsilon-Eta}) of the slow-roll parameters. Following the same procedure we used for the scalar perturbations above, we first rewrite this equation in terms of the conformal time $\tau$ and the new variables $\delta\chi_{ij\mathbf{k}}=a\mathcal{H}_{ij\mathbf{k}}$,
\begin{equation}
\delta\chi''_{ij\mathbf{k}}+\frac{\eta}{2\tau}\delta\chi'_{ij\mathbf{k}}+\left[k^2-\frac{1}{\tau^2}\left(2+3\epsilon-\frac{5\eta}{2}\right)\right]
\delta\chi_{ij\mathbf{k}}=0.
\end{equation}
After solving this equation in the same way as we did for (\ref{SigmaEquation}), we easily deduce the sought-after solution $\mathcal{H}_{ij\mathbf{k}}$,
\begin{equation}\label{TensorModes}
\mathcal{H}_{ij\mathbf{k}}\sim\frac{H}{\sqrt{2k^{3}}}\left(\frac{k}{aH}\right)^{1+\alpha_{t}-\nu_{t}},
\end{equation}
where,
\begin{equation}\label{TensorAlphaNu}
\alpha_{t}=\frac{1}{2}-\frac{\eta}{4}\qquad\mathrm{and}\qquad\nu^2_{t}=\frac{9}{4}+3\epsilon-\frac{11\eta}{4}.
\end{equation}
Having obtained now both the tensor modes' and scalar modes' expressions, we proceed to compute their respective power spectra in the next subsection.
\subsection{The power spectra in the (varying power)-law model}\label{VPLPowers}
Given the solution (\ref{ScalarModes}) we found for the equation of the scalar modes, the power spectrum of the latter is given by \cite{Riotto},
\begin{equation}\label{ScalarPowerSpec}
\mathcal{P}_{\mathcal{R}}=\frac{k^{3}}{2\pi^{2}M_P^2}\frac{H^{2}}{\dot{\phi}^{2}}|\delta\phi_{\mathbf{k}}|^{2}
=\frac{H^{2}}{4\pi^{2}M_P^2\eta^{2}}\left(\frac{k}{aH}\right)^{2+2\alpha_{s}-2\nu_{s}}.
\end{equation}
Here, we have used the definition of our slow-roll parameter $\eta$ in (\ref{Epsilon-Eta}) to obtain the second equality.
Also, given the solution (\ref{TensorModes}) we found for the equation of the tensor modes, the power spectrum of the latter is given by \cite{Riotto},
\begin{equation}\label{TensorPowerSpec}
\mathcal{P}_{h}=\frac{k^{3}}{2\pi^{2}M_P^2}|\mathcal{H}_{ij\mathbf{k}}|^{2}
=\frac{H^{2}}{4\pi^{2}M_P^2}\left(\frac{k}{aH}\right)^{2+2\alpha_{t}-2\nu_{t}}.
\end{equation}
From these two expressions, the tensor-to-scalar ratio $r=\mathcal{P}_{h}/\mathcal{P}_{\mathcal{R}}|_{k=k_*}$ at the Hubble crossing, for which $k_*=a_*H_*$, can immediately be found to be,
\begin{equation}\label{rObtained}
r\simeq\eta_*^2,
\end{equation}
where $\eta_*$ is the slow-roll parameter in (\ref{Epsilon-Eta}) evaluated at the Hubble crossing. The scalar and tensor tilts $n_{s}$ and $n_{t}$ can be computed from the scalar and tensor power spectra $\mathcal{P}_{\mathcal{R}}$ and $\mathcal{P}_{h}$, respectively, as \cite{Riotto},
\begin{equation}\label{ScalarTilt}
n_{s}-1=\frac{d\ln\mathcal{P}_{\mathcal{R}}}{d\ln k}\Bigg|_{k=k_*}=2+2\alpha_{s}-2\nu_{s},
\end{equation}
\begin{equation}\label{TensorTilt}
n_{t}=\frac{d\ln\mathcal{P}_{h}}{d\ln k}\Bigg|_{k=k_{*}}=2+2\alpha_{t}-2\nu_{t}.
\end{equation}

With these expressions in hand, let us now plug in the numerical values and confront them with observations.

First, as we discussed below the definitions (\ref{Epsilon-Eta}) of our slow-roll parameters, two considerations must be taken into account when deciding where to look inside Table \ref{tab:Table}. In fact, we should decide both where to start inflation and where to end it. From the two columns containing the two slow-roll parameters $\epsilon$ and $\eta$, it is clear that we cannot go beyond the energy scale $\rho\sim10^{-0.05}$ without violating the second condition on $\eta$ in (\ref{Epsilon-Eta}).

Next, from the integral (\ref{Efolds2}), we deduce that if inflation has to end around $\rho\sim10^{-0.05}$, it also has to start at most around $\rho\sim10^{-10^{-8}}$ in order to have the desired $N\sim50$. With these boundary conditions, we can compute the corresponding tensor-to-scalar ratio, the scalar tilt and the tensor tilt.

Using the results in Table \ref{tab:Table} we can compute the values of $r$, $n_s$ and $n_t$ in (\ref{rObtained}), (\ref{ScalarTilt}) and (\ref{TensorTilt}), respectively, for various energy scales comprised between $\rho\sim10^{-10^{-8}}$ and $\rho\sim10^{-0.05}$. We easily find that the scale for the Hubble crossing that matches the best with observations is comprised between $\rho\sim10^{-0.008}$ and $\rho\sim10^{-0.01}$ for which the values of $r$, $n_s$, and $n_t$ found using expressions (\ref{ScalarAlphaNu}) and (\ref{TensorAlphaNu}) for $\alpha_{s}$, $\nu_{s}$, $\alpha_{t}$ and $\nu_{t}$ are: $r\in[0.00035,0.00055]$, $n_s\in[0.965,0.972]$, and $n_t\in[0.031,0.040]$,

The specific scale that agrees best with the observed values is found to be $\rho\sim10^{-0.009}$. Indeed for this scale a computation using again expressions (\ref{ScalarAlphaNu}) and (\ref{TensorAlphaNu}) for $\alpha_{s}$, $\nu_{s}$, $\alpha_{t}$ and $\nu_{t}$ gives the following results:
\begin{align}\label{numericals}
r&\simeq 0.00044,\\
n_{s}&\simeq 0.969,\\
n_{t}&\simeq 0.036.
\end{align}

Here we have used the value of the scalar tilt to find the best match. In fact, the value of the ratio $r$ is only known by its upper boundary value which is $r<0.1$ \cite{Planck2015}. Notice that the tensor tilt here is positive in contrast to usual single-field inflationary models where it comes out negative. The other major difference is that the model predicts a very small value for the tensor-to-scalar ratio $r$.

Using (\ref{ScalarPowerSpec}), (\ref{rObtained}), and (\ref{numericals}) we can also deduce the energy scale for inflation required by the model after making use of the relation between the scalar power spectrum $\mathcal{P_{\mathcal{R}}}$ and its amplitude $A_S$ \cite{Riotto}. In fact, the latest observations \cite{Planck2015} show that the amplitude of the scalar power spectrum at the Planck pivot scale $k_*$ is\footnote{The author is grateful to the anonymous referee for having pointed this out.} $A_S\sim2\times10^{-9}$. This implies that the Hubble parameter, i.e., the inflation scale according to our model, should be around $\sim10^{14}\,{\rm GeV}$.

We can see now the effect of having the scalar field in the power of the Ricci scalar. It is thanks to this term that the contribution of the perturbation $\delta R$ appeared on the right-hand side of (\ref{PerturbedKG}) multiplied by the derivative $V_{,\phi R}$. This term could have had a big effect in the spectrum tilt of the scalar modes. But, because the same effect appeared both in the Hubble friction term and the second derivative term of the scalar field the net effect is canceled out and the net result is a very small tilt being proportional to the slow-roll parameters. Also, for the tensor perturbations additional terms appeared in the Hubble friction term but since these terms are due to the slow-roll parameters their effect is very small.
\section{Discussion and conclusion}
We derived in this paper the general dynamical equations for the scalar and tensor modes in any $f(R,\phi)$ modified gravity model. We then applied these equations to the specific (varying power)-law modified gravity model. We have deduced the analogue of the slow-roll parameters in single-field inflationary models proper to this model. We found that the model needs just two slow-roll parameters. By using these parameters and the conditions imposed on them, we were able to identify the adequate \textit{interval} for the energy scale of inflation. We would like to emphasize here that it was only possible to deduce from this model the energy interval for inflation without being able to say anything about the starting point. This being due to the fact that the model is only phenomenological and can only work once an energy scale at which it is to be used has been fixed.

The model allowed us to deduce the power spectra of the perturbations and to extract the tensor-to-scalar ratio, the scalar tilt and the tensor tilt. The numerical values we found for these are in agreement with the latest observations made of the CMB. In fact, we found that the model is able to recover the observed value of the scalar tilt to arbitrary precision. In contrast to usual single-field inflationary models the tensor-to-scalar ratio predicted by the (varying power)-law model is extremely small. Also, the model predicts a positive tensor tilt.

In contrast to what is usually found in usual inflationary models, one does not need to fine tune the parameters of the model in order to recover actual observations. The only free parameters of the model are the scalar field's mass $m$ and the energy scale $\mu$. As the scalar field's mass is not relevant for inflation, the single parameter that was needed was $\mu$. This energy-scale is what determines the energy scale of inflation. By comparing with the latest observations of the density perturbations in the sky, the energy scale required by the model was found to be around $\sim10^{14}\,{\rm GeV}$.

The fact that we needed to have inflation start at a scale that is so close to $\mu$ up to the eighth decimal is not much of a short-coming of the model given that results in agreement with observation could be recovered at an arbitrary precision. The fact that the model requires an inflation that starts at a given energy-scale with such a precision just means that as soon as the Universe reaches that specific energy scale inflation suddenly ignites and lasts until the slow-roll parameters defined above no longer satisfy the necessary conditions to maintain the exponential expansion of the Universe.

Note also, that since this model is phenomenological in nature, it does not provide a theoretical reason for the value of the energy scale it requires for inflation. That scale has been found only after confronting the predictions of the model with observations in the sky.\\

\begin{acknowledgments}
The author is grateful to Patrick Labelle for the very helpful, encouraging and stimulating discussions at various stages of this work, as well as to Khireddine Nouicer for the helpful discussion. The author would also like to thank the anonymous referee for his/her pertinent and very helpful comments that helped improve the manuscript.

This work is supported by the Natural Sciences \& Engineering Research Council of Canada.

\end{acknowledgments}

\end{document}